%% file: output.tex
\documentclass[conference]{IEEEtran}
\makeatletter
\def\ps@headings{%
\def\@oddhead{\mbox{}\scriptsize\rightmark \hfil \thepage}%
\def\@evenhead{\scriptsize\thepage \hfil \leftmark\mbox{}}%
\def\@oddfoot{}%
\def\@evenfoot{}}
\makeatother
\usepackage{cite}
\usepackage{textcomp}
\usepackage{multirow}
\usepackage{graphicx}
\usepackage{subcaption}
\usepackage[most]{tcolorbox}
\usepackage{algorithm,algpseudocode}
\usepackage{optidef}
\usepackage{threeparttable}
\usepackage{makecell}
\usepackage{arydshln}
\usepackage{xspace}
\usepackage{tabularray}
\usepackage{xcolor}
\usepackage{url}
\usepackage{amssymb} 
\usepackage{CJKutf8}
\usepackage[caption=false]{subfig}
\usepackage[T1]{fontenc} 
\usepackage[left=0.63in,right=0.64in,top=0.75in,bottom=1.07in]{geometry}

\newcommand{\be}{\begin{equation}}
\newcommand{\ee}{\end{equation}}

\def\BibTeX{{\rm B\kern-.05em{\sc i\kern-.025em b}\kern-.08em
    T\kern-.1667em\lower.7ex\hbox{E}\kern-.125emX}}

\begin{document}
\bstctlcite{IEEEexample:BSTcontrol}
\newcommand\hxnote[1]{\textcolor{blue}{HX: #1}}
\newcommand\yxnote[1]{\textcolor{magenta}{Yang: #1}}

\title{Closing the Visibility Gap: A Monitoring Framework for Verifiable Open RAN Operations}


\author{
    \IEEEauthorblockN{
        Hexuan Yu\IEEEauthorrefmark{1},
        Md Mohaimin Al Barat\IEEEauthorrefmark{1},
        Yang Xiao\IEEEauthorrefmark{2},
        Y. Thomas Hou\IEEEauthorrefmark{1},
        and Wenjing Lou\IEEEauthorrefmark{1}
    }
    \IEEEauthorblockA{\IEEEauthorrefmark{1}Virginia Polytechnic Institute and State University, VA, USA\\
    \IEEEauthorrefmark{2}University of Kentucky, KY, USA}
}

\maketitle

\begin{abstract}

Open Radio Access Network (Open RAN) is reshaping mobile network architecture by promoting openness, disaggregation, and cross-vendor interoperability. However, this architectural flexibility introduces new security challenges, especially in deployments where multiple mobile network operators (MNOs) jointly operate shared components. 
Existing Zero Trust Architectures (ZTA) in O-RAN, as defined by governmental and industry standards, 
implicitly assume that authenticated components will comply with operational policies. However, this assumption creates a critical blind spot: misconfigured or compromised components can silently violate policies, misuse resources, or corrupt downstream processes (e.g., ML-based RIC xApps).

To address this critical gap, we propose a monitoring framework for low-trust O-RAN environments that proactively verifies configuration state and control behavior against tenant-defined policies. Our system provides scalable, verifiable oversight to enhance transparency and trust in O-RAN operations. We implement and evaluate the framework using standardized O-RAN configurations, with total processing latency of approximately 200 ms, demonstrating its efficiency and practicality for timely policy enforcement and compliance auditing in multi-MNO deployments.
\end{abstract}


\input{1_introduction.tex}

\input{2_background}

\input{3_overview.tex}

\input{4_design}
\input{6_eval.tex}

\input{7_discussion.tex}

\input{9_conclusion.tex}

\bibliographystyle{ieeetr}

\bibliography{reference}

\end{document}

%% file: 1_introduction.tex
\section{Introduction}
\begin{figure*}[]
\centering
\begin{minipage}[t]{0.6\textwidth}
    \vspace{0pt}
    \centering
    \includegraphics[height=4.0cm]{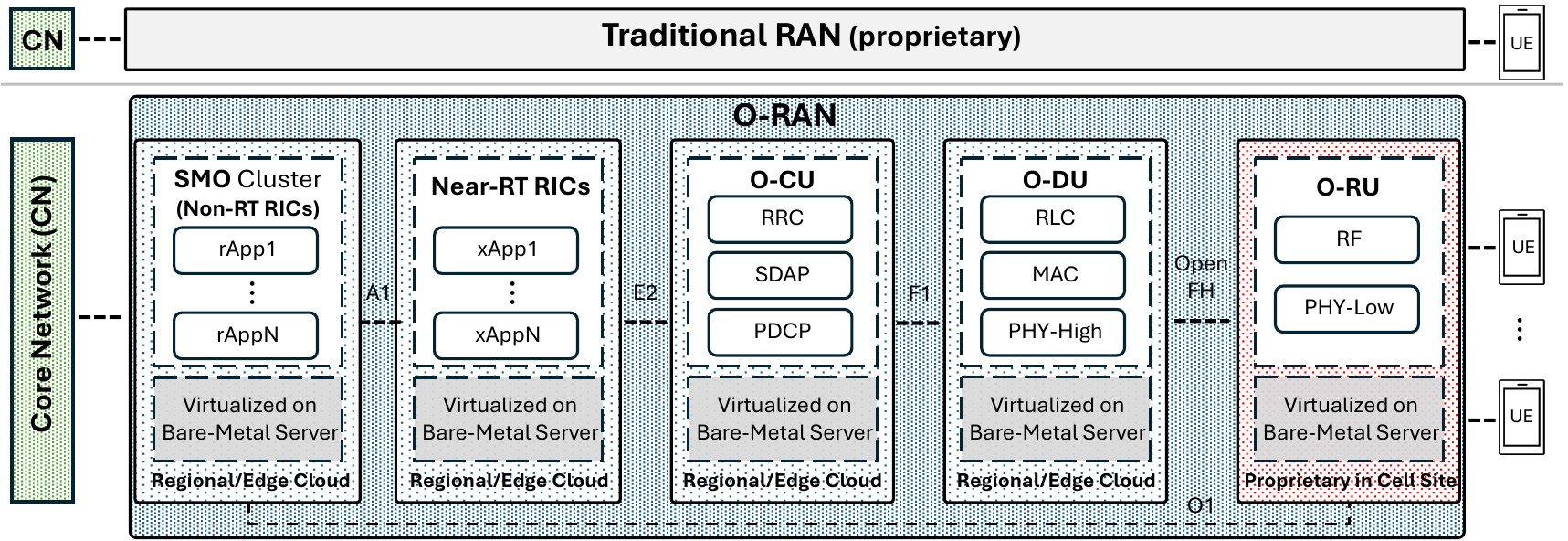}\vspace{-5pt}
    \caption{Traditional monolithic RAN vs. Disaggregated O-RAN architecture}
    \label{fig:oran-architecture}
\end{minipage}
\hfill
\begin{minipage}[t]{0.34\textwidth}
    \vspace{0pt}
    \centering
    \includegraphics[height=3.7 cm]{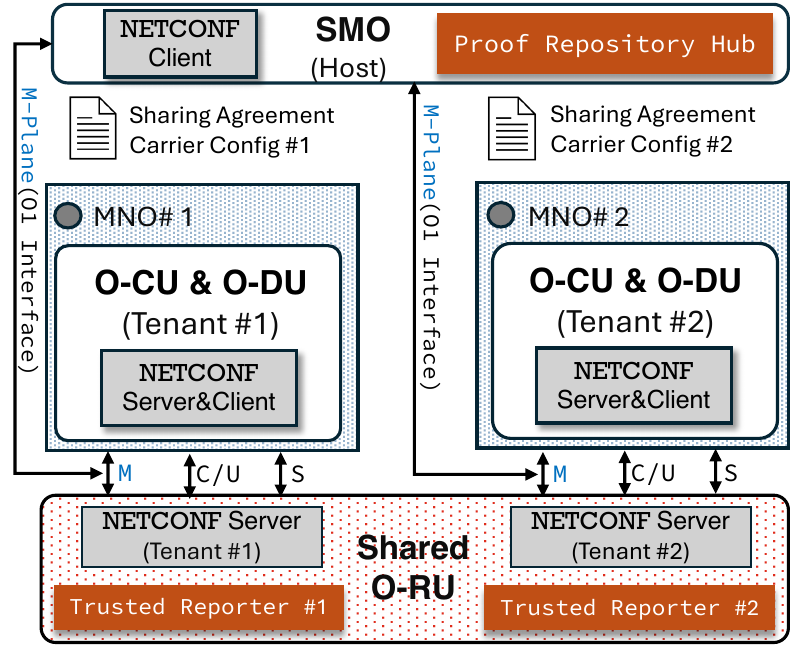}\vspace{-5pt}
\caption{
A shared O-RAN deployment. \\ {\scriptsize *Orange components: our proposed modules \textbf{TR} and \textbf{PRH}.}
}
\label{fig:oru}
\end{minipage}
\vspace{-10pt}
\end{figure*}

Open RAN (specifically O-RAN, referring to the O-RAN Alliance's specifications) is transforming mobile infrastructure by promoting openness and disaggregation. It enables flexible, cost-efficient deployments, reduces vendor lock-in, supports infrastructure sharing, and facilitates AI-driven control for Next-Generation (NextG) mobile networks~\cite{ntiaJointStatement}.\vspace{0.2em}
\noindent \underline{\textbf{O-RAN Architecture.}}
O-RAN decomposes the traditional base station into modular components (illustrated in Fig.~\ref{fig:oran-architecture}): the Radio Unit (\textbf{O-RU}), Distributed Unit (\textbf{O-DU}), and Centralized Unit (\textbf{O-CU}). These components communicate via standardized open interfaces like Open Fronthaul (\textbf{Open FH}). 
Service Management and Orchestration (\textbf{SMO}) and RAN Intelligent Controllers (\textbf{RICs}) provide overarching management and AI/ML-driven control.
ML-driven xApps and rApps, typically deployed as containerized microservices within the RICs, rely on telemetry, particularly Key Performance Measurement (\textbf{KPM}) data, for accurate control decisions such as handovers, traffic steering, and beamforming.\vspace{0.2em}

\noindent \underline{\textbf{Gaps in the Current O-RAN ZTA.}} While disaggregation enhances flexibility, it also broadens the attack surface and complicates trust management, especially in multi-tenant or multi-vendor deployments~\cite{vmware_openran_risks, nec_openran_security}. 
Current ZTA approaches for O-RAN, as guided by industry and governmental standards~\cite{oranzerotrust,nist800207,ORAN2024ThreatModeling}, primarily emphasize identity verification (authentication), access control (authorization), and communication protection (encryption). While effective at boundary protection, they implicitly assume that authenticated components remain compliant at runtime. This assumption creates critical blind spots:

\noindent \textbf{1. Absence of Runtime Compliance Assurance.} Current O-RAN ZTA approaches ensure that only authenticated and authorized components can join the network, but they offer no guarantees about how those components behave after authentication. In practice, authenticated O-RAN components are \emph{implicitly trusted} to operate according to policy. However, without mechanisms for runtime verification, components may silently deviate from their intended configuration due to compromise or misconfiguration, especially when operated by third parties outside a tenant’s administrative control.
    
\noindent \textbf{2. Lack of Tenant-Centric Visibility and Verification.}  In shared deployments, tenants require verifiable assurances that their slice-specific policies (e.g., spectrum use, QoS parameters) are correctly enforced. Yet existing mechanisms provide no way to independently verify whether those policies are being followed or silently violated.

These issues are especially critical at the O-RU~\cite{oran2025sharedoru}, which resides at the cell edge and directly governs RF behavior, synchronization, and fronthaul transmission. Unlike O-DUs and O-CUs, which can be flexibly deployed in operator-managed cloud environments~\cite{bonati2020open,ORAN2020usecases}, O-RUs are typically vendor-controlled and expose only partially standardized, limited-function interfaces. 
They are managed remotely over the management (\textbf{M-Plane}) and control (\textbf{C-Plane}) planes by third-party operators or neutral hosts. This limits tenant visibility into the O-RU’s configuration and behavior, increasing the risk of undetected faults or malicious deviations. Even an authenticated O-RU can silently violate policies, falsify telemetry (e.g., KPMs), or mislead downstream control decisions. 
This opaque trust model is especially concerning as O-RAN becomes central to mission-critical applications like dynamic spectrum sharing, industrial automation, and AI-driven orchestration~\cite{pinson2025openran, ntiaAdvancedDynamic,gopal2024prosas,smith2021ran}.

These concerns are amplified by weak protections on the fronthaul interfaces. Historically, 5G fronthaul standards made integrity protection \textit{optional}, assuming that link-layer attacks would be unlikely in a controlled environment. However, recent research has shown that this assumption is problematic in O-RAN, where the protection remains optional and inconsistently deployed: 
 when left unprotected, adversaries can launch man-in-the-middle (MITM) attacks and manipulate fronthaul traffic to cause silent RF misbehavior, such as manipulating beamforming parameters or injecting malformed IQ samples~\cite{xing2024criticality, abdalla2024end}.
Yang et al.~\cite{yang2024oranalyst} revealed widespread vulnerabilities due to insecure M-plane configurations in real-world O-RAN deployments. 
Akon et al.~\cite{akon2025control} demonstrated weak 5G access control and insufficient isolation configurations enable stealthy privilege escalation and cross-tenant attacks. 
An NTIA report~\cite{NTIA2023OpenRANSecurity} similarly highlights M-plane exposure to misconfiguration, leakage, and control abuse.
Tabiban et al.~\cite{tabiban2023signaling} showed that weak protection on open FH can enable adversaries to impersonate O-DUs and trigger C-plane disruptions such as spoofed handovers and signaling overloads. 
Additional works~\cite{maamary2024synchronization,baguer2024attacking} identified that the synchronization plane (\textbf{S-Plane}) in O-RAN is also vulnerable to configuration attacks, such as Precision Time Protocol (PTP) parameter tampering and timing spoofing.

Misconfigurations, whether accidental or malicious, can disrupt higher-layer control functions by introducing malformed inputs to components like the RICs and degrading the behavior of dependent xApps and rApps~\cite{yungaicela2024misconfiguration}. 
In particular, corrupted telemetry data, such as KPMs and their derived Key Performance Indicators (KPIs)
, can mislead decision logic across the stack. For example, poisoned KPM data can degrade ML model accuracy~\cite{soltani2025intelligent, chiejina2024system}. In targeted attacks, malicious cells can report manipulated KPIs to deceive traffic steering mechanisms, unfairly attract User Equipment (UE) connections, and violate QoS or fairness constraints~\cite{aizikovich2025rogue}.

Collectively, these findings expose a fundamental limitation in current O-RAN deployments: while tenants can declare policy goals, e.g., enabling encryption, defining RF constraints, or setting telemetry rules, they lack any assurance that these configurations are correctly enforced or remain intact at runtime. The issue is not the absence of security mechanisms, but the inability to validate their operational status, such as whether fronthaul integrity protection is active, among other status items, especially in shared, partially trusted infrastructure where management is delegated to third-party or neutral-host.\vspace{0.2em}

\noindent \underline{\textbf{Our Solutions.}} To address the lack of runtime compliance assurance and tenant-verifiable behavior in O-RAN deployments, we design a monitoring framework that shifts trust from static credentials to operational evidence. Our goal is not to prevent specific attacks directly, but to restore timely and verifiable visibility in environments where O-RAN components may be shared, remotely managed, or operated by untrusted parties. Existing ZTA authenticates entities at the boundary but assumes compliant behavior post-authentication. We extend this model with continuous, timely, and hardware-rooted verification of configuration and control state, enabling tenants to audit and respond to policy deviations throughout the O-RAN lifecycle.\vspace{0.2em}
\noindent \textit{Design Overview.} 
Our monitoring framework comprises two core components (Fig.~\ref{fig:oru}): the \textbf{Trusted Reporter (TR)}, deployed within the O-RU controller, and the \textbf{Proof Repository Hub (PRH)}, deployed as an rApp at the SMO layer. 
The security of both components is anchored by a hardware root of trust, typically a Trusted Platform Module (TPM), a standardized crypto-processor that provides a hardware root of trust and is widely available on RAN devices, as mandated by O-RAN~\cite{oran2022ocloud, oranzerotrust} and 3GPP standards~\cite{3gpp-ts-33.501}, which promote trusted hardware (e.g., TPM, TEE) to isolate sensitive processes and data. This ensures that monitoring and reporting remain tamper-resistant, even in the presence of a compromised host, serving as the only trusted anchors in an otherwise untrusted infrastructure.

The TR monitors management and control interfaces, capturing critical events (e.g., configuration changes) that may affect radio behavior. 
To minimize latency impact, it performs no local enforcement; instead, it generates tamper-evident, cryptographically signed logs capturing critical states and observed configurations. 
For efficiency, logs are organized into a Merkle tree, supporting compact representation and incremental root updates. The resulting root is signed and transmitted as a single report to the PRH, which maintains an immutable audit trail and empowers tenants to detect policy violations and initiate mitigation via the SMO.
Our prototype, evaluated with standard O-RAN configurations, demonstrates total processing latency of approximately 200 ms per monitoring cycle, affirming the framework’s practicality for real-world policy enforcement. 
The \emph{contributions} of this paper can be summarized as follows:\\
\noindent \textbf{1. Runtime Monitoring for O-RAN:} We present a monitoring framework anchored in a hardware root of trust that attests O-RU configuration and operational state, enabling continuous compliance assurance in shared or federated deployments.\\
\noindent \textbf{2. Scalable, Tenant-Verifiable Evidence Reporting:} We provide tenants with Merkle tree-backed reports that offer compact, verifiable snapshots of the operational state. This enables tenants to respond promptly to policy violations before misconfigurations or malicious actions can propagate.\\
\noindent \textbf{3. Contextual Trust Extension for ZTA:} 
We bridge the gap between static identity verification and dynamic policy compliance by continuously validating control and configuration state, aligning Zero Trust principles with operational enforcement.

%% file: 2_background.tex
\section{Background and Related Work}
\subsection{Background of O-RAN} 
O-RAN introduces a modular architecture with standardized interfaces, enabling vendor-neutral deployment of components like the O-RU, O-DU, O-CU, and RICs. 
Among these, the O-RU is deployed at the cell site and handles time-sensitive RF and lower PHY functions (e.g., beamforming and frame synchronization) through specialized hardware (e.g., FPGAs, ASICs, DSPs, and RF transceivers). Its operation is governed by a vendor-supplied embedded software stack, referred to as the \emph{O-RU controller}~\cite{etsi104023}, typically implemented over embedded Linux or real-time operating system (RTOS) to enable remote management and secure execution.
 The O-RU’s operation is structured across four logical planes (Fig.~\ref{fig:oru}):\\
\noindent \underline{\textbf{1. M-Plane:}} Manages configuration and lifecycle tasks via the XML-based Network Configuration Protocol (\textbf{NETCONF})~\cite{rfc6241,etsi104023} over secure channels like SSH/TLS. NETCONF defines a very large portion of an O-RU's configurable behavior (its policies, features, interfaces, security parameters, etc.) and follows a client-server model via Remote Procedure Calls (RPCs), using YANG~\cite{rfc6020} as the data modeling language to define configuration and state data structures. The SMO acts as the client, while the O-RU controller exposes server endpoints through a programmable interface. In some cases, the O-DU may also operate as a client toward the O-RU.  This setup enables the SMO to configure RF parameters (e.g., carrier frequency, TX/RX gains), manage software updates and synchronization, and monitor device health.
Example functions include O-RU identification (e.g., via \texttt{o-ran-dhcp} YANG model) and X.509 certificate provisioning for TLS connections.\\
\noindent \underline{\textbf{2. C-Plane:}} Delivers real-time control information (e.g., beamforming coefficients, scheduling commands) from the O-DU over the Enhanced Common Public Radio Interface (eCPRI) protocol~\cite{ecpri2017}. The O-RU can monitor the C/U connection health using internal watchdog timers, e.g., the \texttt{o-ran-supervision.yang} model.\\
\noindent \underline{\textbf{3. U-Plane:}} Transports data traffic (e.g., IQ samples) over fronthaul links between the O-RU and O-DU.\\
\noindent \underline{\textbf{4. S-Plane:}} Maintains precise timing and frequency synchronization, typically using \texttt{IEEE 1588} PTP~\cite{ieee1588}. Key parameters (e.g., \texttt{clock-class}) are exposed via the M-Plane using models like \texttt{o-ran-ptp.yang}.

\noindent Other core components include:\\
\noindent \underline{\textbf{O-DU:}} Implements Radio Link Control (RLC), Medium Access Control (MAC), and High-PHY layers. It manages scheduling, mobility, and QoS, interfacing with the O-RU over Open FH and with the O-CU over the F1 interface.
It also runs both NETCONF client and server roles, depending on the interaction target.\\
\noindent \underline{\textbf{O-CU:}} Operates the Packet Data Convergence Protocol (PDCP), Service Data Adaptation Protocol (SDAP), and Radio Resource Control (RRC) layers, managing signaling and encrypted data transfer to the core network (CN).\\
\noindent \underline{\textbf{SMO} \& \textbf{Non-Real-Time RIC (Non-RT RIC)}}: The SMO orchestrates RAN operations, including configuration and monitoring of O-RUs and O-DUs via NETCONF RPCs. Its integrated Non-RT RIC enforces policies and manages ML workflows to support the Near-RT RIC.\\
\noindent \underline{\textbf{Near-Real-Time RIC (Near-RT RIC)}}: Executes xApps for real-time RAN control, such as handover and load balancing, using telemetry from the E2 interface.

\vspace{-4pt}
\subsection{Related Work}
As security and privacy of network communication systems evolve~\cite{du2023ucblocker, yuaaka, du2022mobile, li2023bijack, zhang2023mindfl, zhang2024hermes, wang2025feco, zhang2024state, zhang2025starcast, zhang2025sentinel}, securing O-RAN deployments has become an active research area due to its expanded attack surface. A substantial body of work applies AI/ML techniques to detect anomalous behavior based on monitored telemetry. Xavier et al.~\cite{10279349} and Scalingi et al.~\cite{scalingi2024det} use supervised and data-driven models to detect early-stage control-plane attacks.  Spotlight~\cite{sun2024spotlight} improves detection accuracy and explainability in xApp-based telemetry analysis. 
Aizikovich et al.~\cite{aizikovich2025rogue} investigate adversarial telemetry manipulation targeting traffic steering mechanisms and design a learning-based defense. Tabiban et al.~\cite{tabiban2023signaling} and Wen et al.~\cite{wen20245g} focus on signaling-layer threats, proposing learning based detection via xApps/rApps to watch for anomalies in control messages. Additionally, several efforts focus on improving scheme efficiency~\cite{yu2021fpga, zhang2021high, fan2019gpu}. While these methods are effective for post hoc anomaly detection, they depend on the integrity of upstream telemetry and are inherently reactive, relying on statistical heuristics. In contrast, our system focuses on generating verifiable evidence of configuration and control actions at the source, narrowing the gap between observed behavior and provable ground truth.

Besides, Foukas et al.~\cite{foukas2023taking} introduce a programmable RAN monitoring architecture using eBPF to inject safe codelets for performance telemetry and dynamic control. Although effective for optimization, their approach targets operational efficiency rather than adversarial threats. 
Atalay et al.~\cite{atalay2024openran} propose a TEE-based authentication and authorization system for xApps, isolating critical control logic from untrusted workloads. ZTRAN~\cite{abdalla2024ztran} applies Zero Trust principles in the RIC by deploying xApps that enforce access control across management services. Similarly, Balasingam et al.~\cite{balasingam2024application} present a service-level assurance framework for RAN slicing, ensuring application-layer reliability but not infrastructure-layer integrity. These efforts secure individual components or services but do not address the verifiability of low-level, runtime behavior in shared or edge-deployed components, e.g., O-RU.

%% file: 3_overview.tex
\section{System and Threat Model}
\subsection{System Model}

We design a monitoring and verification framework for O-RAN deployments involving low-trust or shared infrastructure (Fig.~\ref{fig:oru}), where policy violations may arise from compromised or misbehaving components (e.g., NETCONF clients) deployed across disparate trust domains. It focuses on protecting the configuration and control surfaces exposed over the M-plane, C-plane, and S-plane. These planes, programmable through standardized interfaces such as NETCONF/YANG, are commonly targeted in misconfiguration, policy violation, or insider attack scenarios  (e.g., unauthorized RF changes, beamforming directives, or altered PTP settings).

To establish operational assurance, each tenant-O-RU instance is paired with a logically isolated \textbf{TR}, whose signing and attestation are bound to the TPM to securely monitor and report the behavior of the O-RU controller.
The TR records configuration and control operations issued to the O-RU via the M-,C-, and S-planes, either periodically or event-triggered (see Section~\ref{trigger}). 
These observations are aggregated into a Merkle tree, with the resulting 256-bit root serving as a compact snapshot of the system state. The TR produces a signed attestation report over this root, which is delivered to a tenant-visible \textbf{PRH}, hosted within cloud-based TPMs (or TEEs) at the SMO layer for evidence logging and policy verification by tenants.  If violations are detected, e.g., disabled encryption, tenants can initiate remediation actions via the SMO, e.g., revoking access or updating policies.

\subsection{Threat Model}
We consider a threat model where O-RAN components such as the O-RU, O-DU, O-CU, and SMO may be operated by third-party vendors or span across distinct administrative domains. These components can be unintentionally misconfigured, operated by untrusted parties, or deliberately compromised by external adversaries. As such, we do not assume trust in the host OS or hypervisor. Instead, we assume trust in the hardware-protected attestation keys and associated measurement mechanisms, which together form the system’s \textit{root of trust}. These may be provided by a TPM, a TEE, or a combination of both. The \textit{authenticity} of the reports is verified via cryptographic attestation signatures, while platform measurements provide assurance that the correct monitoring program was loaded (i.e.,integrity). We define platform measurements as cryptographic hashes representing either the system’s boot or runtime state (e.g., TPM Platform Configuration Registers (PCRs)) or a hardware-reported hash of a secure execution environment (e.g., \texttt{MRENCLAVE} in Intel SGX).

Our framework targets threats to the M-, C-, and S-planes, which expose programmable surfaces through open interfaces. These planes are frequent vectors for misconfiguration, policy violations (e.g., scheduling manipulation, disabled synchronization), and external threats (e.g., MITM). We do not inspect or intervene in the U-plane, which carries high-throughput user traffic and is protected through independent 3GPP mechanisms like PDCP encryption. Instead, we verify the correctness of upstream configuration and control logic that governs U-plane behavior, and ensure that interfaces (e.g., open FH, O1) are properly secured to prevent unauthorized changes that could indirectly impact radio operations or information leakage.

We exclude physical and hardware-level attacks from our scope. These include FPGA tampering, RF circuitry modifications, side-channel exploits, and memory bus probing, etc., which require invasive tools and physical access. Such attacks are typically mitigated by secure boot, firmware signing, and JTAG lockdowns, as mandated by O-RAN specifications~\cite{oranWG11security}. Although the O-RU executes lower-layer PHY functions in hardware (e.g., FPGAs, ASICs), its operational logic is dictated by externally issued instructions. These runtime instructions, rather than static firmware or logic, represent the dominant attack surface in modern O-RAN deployments and form the focus of our monitoring framework.

%% file: 4_design.tex
\section{Design Details}
This section presents the design of our monitoring and verification framework, which enforces runtime policy compliance and enables verifiable oversight of configuration and operational behavior across the M-, C-, and S-planes. The system emphasizes secure evidence collection and supports efficient validation using Merkle tree–based cryptographic reporting.

\begin{table*}[t]
\centering
\small
\caption{Selected Critical Evidence Items to be Monitored}\vspace{-5pt}
\resizebox{0.8\textwidth}{!}{%
\begin{tabular}{|p{3.95cm}|p{5.4cm}|p{6.2cm}|}
\hline
\textbf{Category} & \textbf{Security Relevance} & \textbf{Attestation Method} \\
\hline
Runtime Configuration State\textsuperscript{*} & Determines real-time RF, telemetry, and security behavior; deviations can silently violate tenant policies & Hash critical subtrees from \texttt{<running>} datastore using models like \texttt{o-ran-ru}, \texttt{o-ran-kpm}, \texttt{ietf-macsec}, \texttt{o-ran-ptp} \\
\hline
Startup Configuration Baseline & Serves as the intended configuration reference; divergence indicates runtime policy violations or drift & Compare \texttt{startup-config.xml} with current operational state and hash for audit purposes \\
\hline
Trusted Reporter Binary & Verifies the integrity of the Trusted Reporter code & Verify attestation report containing platform measurement (e.g., enclave or binary hash) against expected build-time hash\\
\hline
\mbox{Access Control \& Permissions}\textsuperscript{*} & Prevents privilege escalation via misconfigured access rules for users and processes & Hash user roles and RPC permissions defined in \texttt{o-ran-usermgmt} and \texttt{ietf-netconf-acm} subtrees \\
\hline
\mbox{KPM Telemetry Configuration}\textsuperscript{*} & Controls granularity and content of reports used by RIC/xApps & Hash selected fields from \texttt{o-ran-kpm} (e.g., measurement objects, reporting interval) \\
\hline
KPM Report Subscription\textsuperscript{*} & Prevents misdirection or disabling of telemetry reports & Hash target endpoint and subscription path in \texttt{o-ran-kpm} \\
\hline
\mbox{Fronthaul Security}\textsuperscript{*} & Validates MACsec/IPsec settings to prevent MITM attacks on eCPRI & Hash fields from \texttt{ietf-macsec} or \texttt{ietf-ipsec} (e.g., \texttt{macsec-enabled}) \\
\hline
\mbox{Synchronization Parameters}\textsuperscript{*} & Misconfigured timing causes service degradation or DoS & Attest fields like \texttt{clock-class}, \texttt{domain-number} from \texttt{o-ran-ptp} or \texttt{ietf-ptp} \\
\hline
Firmware / Software Version\textsuperscript{*} & Prevents rollback or unpatched binaries & Hash version string from \texttt{o-ran-ru} or verify firmware metadata if accessible \\
\hline
RF Capabilities Declaration\textsuperscript{*} & Prevents false claims of supported features & Hash fields from \texttt{o-ran-ru} YANG model \\
\hline
YANG Module Integrity & Prevents schema spoofing or stealth model manipulation & Hash installed model files (e.g., \texttt{/etc/sysrepo/yang/*.yang}) \\
\hline
Init and System Services & Prevents unauthorized control-plane logic injection & Hash trusted init scripts or unit files (e.g., \texttt{/etc/systemd/system/}) \\
\hline
\end{tabular}
}
\label{tab:oru-attestation-scope}\vspace{-3pt}
\begin{flushleft}
\textsuperscript{*} \textit{Indicates evidence derived from YANG-modeled NETCONF datastores, typically retrieved at runtime via NETCONF or extracted from persisted NETCONF snapshots. Unmarked items are verified via local file system access or attestation-based integrity reporting.}
\end{flushleft}\vspace{-15pt}
\end{table*}

\subsection{Monitoring Scope and Methods}

Our framework strategically monitors configuration and high-impact runtime targets to establish trust in the O-RU’s behavior. 

Table \ref{tab:oru-attestation-scope} presents a selection of these critical targets, detailing their security relevance and the attestation methods.

\subsubsection{M-plane Monitoring}
The M-plane is central to defining O-RU behavior, with some key aspects including:\\
\noindent \textbf{NETCONF Runtime Validation.}
We monitor the O-RU's M-Plane, specifically the NETCONF server and its associated YANG models (e.g., \texttt{o-ran-ru, o-ran-kpm}). The TR within the O-RU controller attests the running datastore, hashing critical subtrees (Table~\ref{tab:attestation-scope-summary}) to detect potential unauthorized changes, policy deviations, or configuration drift from the \texttt{startup-config.xml} baseline.\\
\noindent \textbf{Permission Compliance.} 
The TR monitors the configuration of access control mechanisms such as \texttt{ietf-netconf-acm}, \texttt{ietf-netconf-server}, and \texttt{o-ran-usermgmt}.
It verifies that secure transport (e.g., SSH/TLS) is enforced, that access control policies are syntactically valid, and that overly permissive patterns (e.g., wildcard RPC permissions) are absent.  We also attest that the O-RU’s local identity, including certificates and keys established via mechanisms like O-RAN’s certificate management procedures (e.g., CMPv2~\cite{etsi104023}), is correctly provisioned and actively enforced. 

\noindent \textbf{Fronthaul Security Enforcement.} The TR explicitly attests the M-Plane configuration of MACsec (\texttt{ietf-macsec}) or IPsec (\texttt{ietf-ipsec}) to ensure that data transiting this link is protected against MITM attacks. For example, this can be checked through \texttt{macsec-enabled} status, active \texttt{macsec/security-association} entries, and \texttt{ipsec/sa} and \texttt{ipsec/policy} fields when IPsec is used. This enables tenants to confirm cryptographic protections are properly configured and enforced on fronthaul links.\\
\noindent \textbf{Interface Configuration.}
The O-RU’s logical and physical interfaces are defined by the \texttt{o-ran-interfaces.yang} module, covering Ethernet ports, VLANs, and access control mechanisms such as \texttt{IEEE 802.1X}. The TR monitors changes to these settings to detect violations, such as unauthorized ports activation or disabled protections.
\subsubsection{C-plane Monitoring}
To capture dynamic radio behavior, we monitor selected runtime states on the C-plane, which handles low-latency fronthaul control (e.g., via eCPRI).  
We focus on attesting high-impact runtime metadata that govern radio behavior, e.g., \texttt{RU Port ID}, \texttt{BandSector ID}, \texttt{Beam IDs}.
The O-RU controller exports these fields via secure memory interfaces provisioned by the O-RU software and passes them to the TR, where they are bound to platform's measurement chain and the signed evidence logs.  Besides, we adopt delta-based reporting, which captures only control-state changes and minimizes overhead. This ensures that critical operations remain auditable and policy-compliant without compromising the real-time requirements of the C-plane.  For broader coverage, optional RTOS attestation mechanisms (e.g.,~\cite{surminski2021realswatt,wang2023ari}) can be integrated with predictable latency.

\subsubsection{S-Plane Monitoring}
Synchronization enforcement in O-RAN spans both M-plane (via NETCONF) and S-plane (via eCPRI).  
Static parameters such as \texttt{clock-class}, \texttt{domain-number}, and \texttt{priority} are extracted from NETCONF datastores using standardized YANG models (e.g., \texttt{o-ran-ptp}, \texttt{ietf-ptp}) and attested using the same approaches we described for M-plane.
For dynamic behavior, we reuse the C-plane monitoring strategy, the TR records sync indicators, e.g., timestamp offset and clock adjustment logs, via secure memory or telemetry interfaces, and incorporates them into the attested evidence stream. This layered approach ensures both configuration correctness and secure synchronization.

\subsection{Workflow and Design Logic}
The TR, co-located with each O-RU controller, collects a set of evidence items, such as configuration states, synchronization parameters, and software integrity metadata, as described previously. 
Each item is hashed and aggregated into a Merkle tree (Algorithms~\ref{alg:merkle-build},\ref{alg:merkle-update}), and the root is signed along with a timestamp, nonce, and platform identity (Algorithm~\ref{alg:report-generation}) using a key protected by the trusted hardware (e.g., TEE report key, TPM's attestation key). This key remains inaccessible to the untrusted O-RU software stack, ensuring that only a provisioned and verifiable device can produce valid reports. The signed attestation report is transmitted to the PRH, which securely stores the report (e.g., via TEE) and enables tenants to perform verification (Section~\ref{verify}), and take actions if violations are detected. This method guarantees that configuration changes are both authorized and tamper-evident, with strong guarantees of authenticity, integrity, and non-repudiation.

\subsubsection{Replay Protection} To ensure freshness and prevent replay attacks, each report includes a timestamp $t$ derived from a trusted time source by the underlying hardware root-of-trust, as shown in Algorithm~\ref{alg:report-generation}. Unlike PTP-based synchronization used for RF coordination, this timestamp is resistant to tampering by the host OS or hypervisor.  While the specific mechanism varies across platforms, most trusted hardware provide such trusted time sources, e.g.,  \texttt{sgx\_get\_trusted\_time()} in Intel SGX, or \texttt{TPMS\_CLOCK\_INFO} in TPM 2.0. Our framework is hardware-agnostic and only assumes the availability of a cryptographically verifiable time or counter, ensuring each report reflects a unique, temporally anchored system state.


\begin{algorithm}[]
\small
\caption{Tamper-Evident Report Generation in \textbf{TR}}
\begin{algorithmic}[1]
\State $\mathcal{E} \gets \texttt{collect\_evidence\_items}()$
\State $r, \mathcal{T} \gets \textsc{BuildMerkleTree}(\mathcal{E})$ \Comment{\scriptsize{$r$: Merkle root, $\mathcal{T}$: tree structure}}
\State $t \gets \texttt{current\_time}()$ \Comment{HW-secured timestamp}
\State $\texttt{platform\_id} \gets \texttt{get\_tee\_measurement}()$
\State $\texttt{nonce} \gets \texttt{generate\_nonce}()$

\State $m \gets r \,\|\, t \,\|\, \texttt{nonce} \,\|\, \texttt{platform\_id}$ \Comment{Report message to bind}
\State $h_{\text{report}} \gets H(m)$
\State $\sigma \gets \texttt{sign\_with\_attest\_key}(h_{\text{report}})$

\State \Return \texttt{report}:
\State \hspace{1em} $\{$
\State \hspace{2em} \texttt{merkle\_root}: $r$,
\State \hspace{2em} \texttt{timestamp}: $t$,
\State \hspace{2em} \texttt{nonce}: \texttt{nonce},
\State \hspace{2em} \texttt{platform\_id}: \texttt{platform\_id},
\State \hspace{2em} \texttt{signature}: $\sigma$
\State \hspace{1em} $\}$
\end{algorithmic}
\label{alg:report-generation}
\end{algorithm}

\vspace{-8pt}
\subsubsection{Periodic and Event Driven Attestation}\label{trigger}
Attestation is triggered by both periodic timers and \texttt{config-update} events, as described in Algorithm~\ref{alg:attestation-trigger}. The TR recomputes hashes either upon configuration changes or at regular intervals (e.g., every 300 ms, configurable via the \texttt{o-ran-supervision.yang} model~\cite{etsi104023}). This hybrid strategy balances measurement freshness with runtime efficiency.

\begin{algorithm}[]
\small
\caption{Attestation Trigger Logic (\scriptsize{Periodic and Event-Driven})}
\begin{algorithmic}[1]
\State $\texttt{last\_attested\_time} \gets 0$ \Comment{\scriptsize{Initialize last trigger timestamp}}
\State $\texttt{last\_attested\_hash} \gets \perp$ \Comment{Initialize last known config hash}
\While{true}
    \State $\texttt{now} \gets \texttt{current\_time}()$ \Comment{Get current time}
    
    \If{$\texttt{now} - \texttt{last\_attested\_time} \geq T$} \Comment{Periodic trigger}
        \State $\texttt{current\_hash} \gets \texttt{compute\_config\_hash}()$
        \If{$\texttt{current\_hash} \neq \texttt{last\_attested\_hash}$}
            \State $\texttt{report} \gets \texttt{generate\_attestation\_report}()$
            \State \texttt{send\_to\_PRH(report)}
            \State $\texttt{last\_attested\_hash} \gets \texttt{current\_hash}$
            \State $\texttt{last\_attested\_time} \gets \texttt{now}$
        \EndIf
    \EndIf

    \If{\texttt{received\_config\_update\_event}()} \Comment{Event-driven trigger}
        \State $\texttt{current\_hash} \gets \texttt{compute\_config\_hash}()$
        \If{$\texttt{current\_hash} \neq \texttt{last\_attested\_hash}$}
            \State $\texttt{report} \gets \texttt{generate\_attestation\_report}()$
            \State \texttt{send\_to\_PRH(report)}
            \State $\texttt{last\_attested\_hash} \gets \texttt{current\_hash}$
            \State $\texttt{last\_attested\_time} \gets \texttt{now}$
        \EndIf
    \EndIf
\EndWhile
\end{algorithmic}
\label{alg:attestation-trigger}
\end{algorithm}

\subsubsection{Scalable Evidence Reporting via Merkle Trees}
\begin{figure}[t]
\centering
\small
\resizebox{0.7\columnwidth}{!}{
\begin{tikzpicture}[
    level 1/.style={sibling distance=4cm},
    level 2/.style={sibling distance=1.8cm},
    every node/.style={align=center},
    hash/.style={rectangle, draw, rounded corners, minimum width=1.3cm, minimum height=0.3cm},
    edge from parent/.style={draw, thick}
    ]
\node[hash] (root) {\scriptsize{$H(h_1 \| h_2)$}}
  child {node[hash] (H1) {\scriptsize{$h_1 = H(\texttt{ru-cap} \| \texttt{macsec})$}}
    child {node[hash] (L1) {\scriptsize{\texttt{ru-cap}}}}
    child {node[hash] (L2) {\scriptsize{\texttt{macsec}}}}
  }
  child {node[hash] (H2) {\scriptsize{$h_2 = H(\texttt{ptp.cfg} \| \texttt{tr-bin})$}}
    child {node[hash] (L3) {\scriptsize{\texttt{ptp.cfg}}}}
    child {node[hash] (L4) {\scriptsize{\texttt{tr-bin}}}}
  };
\end{tikzpicture}\vspace{-2pt}}
\caption{A simple Merkle tree constructed over 4 evidence items. Internal nodes $h_1$, $h_2$ represent hashes of paired leaves, and the root $H(h_1 \| h_2)$ is signed by the TR. \small (* \texttt{tr-bin} = TR binary)}
\label{fig:merkle-tree}\vspace{-10pt}
\end{figure}
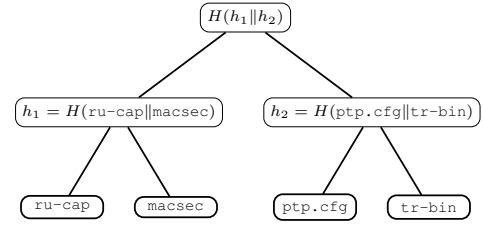

To enable scalable attestation, the TR constructs a Merkle tree~\cite{crosby2009efficient} over a fixed set of evidence items. Each item is hashed into a leaf using SHA-256, and internal nodes are recursively computed as $h = \texttt{SHA256}(h_{\text{left}} \,\|\, h_{\text{right}})$, until the Merkle root is obtained. As discussed previously, the final Merkle root is signed by the trusted hardware along with a timestamp, nonce, and platform identity (Algorithm~\ref{alg:report-generation}) to produce a tamper-evident report. Figure~\ref{fig:merkle-tree} illustrates a simple tree built from representative evidence items such as \texttt{/macsec-enabled} and the TR binary. This design compactly aggregates heterogeneous evidence into a fixed-size digest, enabling lightweight attestation.

Another key advantage of this structure is its support for efficient incremental updates. When only a subset of evidence items changes, the TR recomputes just the affected leaf hashes and their paths to the root (Algorithm~\ref{alg:merkle-update}), avoiding full-tree recomputation and reducing overhead. This design ensures that runtime reporting remains computationally efficient as the number of monitored items scales, while maintaining compact report payloads suitable for frequent policy enforcement in large-scale O-RAN deployments.
Besides, it can be extended to support Merkle proof-based validation~\cite{merkle1987digital}, enabling tenants to verify specific claims, e.g., \texttt{macsec-enabled=true}, thus supporting lightweight and fine-grained verification.

\begin{algorithm}[]
\small
\caption{Merkle Tree Construction in \textbf{TR}}
\begin{algorithmic}[1]
\Function{BuildMerkleTree}{$\mathcal{E}$}
    \Comment{$\mathcal{E} = [e_1, e_2, \dots, e_n]$ is the set of evidence items}
    \State $\mathcal{L}_0 \gets [\texttt{SHA256}(e_i) \text{ for each } e_i \in \mathcal{E}]$ \Comment{\scriptsize{Hash leaf nodes}}
    \State $\mathcal{T} \gets [\mathcal{L}_0]$ \Comment{Initialize Merkle tree structure}
    \While{$|\mathcal{L}_0| > 1$}
        \State $\mathcal{L}_1 \gets [\,]$
        \For{$i = 0$ to $|\mathcal{L}_0| - 1$ by $2$}
            \If{$i + 1 < |\mathcal{L}_0|$}
                \State $h \gets \texttt{SHA256}(\mathcal{L}_0[i] \,\|\, \mathcal{L}_0[i+1])$
            \Else
                \State $h \gets \texttt{SHA256}(\mathcal{L}_0[i] \,\|\, \mathcal{L}_0[i])$ \Comment{Duplicate if odd}
            \EndIf
            \State $\mathcal{L}_1.\texttt{append}(h)$
        \EndFor
        \State $\mathcal{T}.\texttt{append}(\mathcal{L}_1)$
        \State $\mathcal{L}_0 \gets \mathcal{L}_1$
    \EndWhile
    \State \Return $\mathcal{T}[-1][0], \mathcal{T}$ \Comment{Return Merkle root and tree}
\EndFunction
\end{algorithmic}
\label{alg:merkle-build}
\end{algorithm}
\begin{algorithm}[]
\small
\caption{Merkle Tree Incremental Update in \textbf{TR}}
\begin{algorithmic}[1]
\Function{UpdateMerkleTree}{$\mathcal{T}, j, e_j'$}
    \Comment{\scriptsize{Update leaf $j$ with new evidence $e_j'$}}
    \State $\mathcal{T}[0][j] \gets \texttt{SHA256}(e_j')$ \Comment{Update affected leaf}
    \State $current \gets j$
    \For{$\ell = 1$ to $|\mathcal{T}| - 1$}
        \State $p \gets \lfloor current / 2 \rfloor$
        \State $left \gets \mathcal{T}[\ell-1][2p]$
        \If{$2p + 1 < |\mathcal{T}[\ell-1]|$}
            \State $right \gets \mathcal{T}[\ell-1][2p + 1]$
        \Else
            \State $right \gets left$ \Comment{Duplicate if no right sibling}
        \EndIf
        \State $\mathcal{T}[\ell][p] \gets \texttt{SHA256}(left \,\|\, right)$
        \State $current \gets p$
    \EndFor
    \State \Return $\mathcal{T}[-1][0]$ \Comment{Return updated Merkle root}
\EndFunction
\end{algorithmic}
\label{alg:merkle-update}
\end{algorithm}

\subsubsection{Report Verification and Response}\label{verify}
Once the signed attestation report is transmitted to the PRH, the tenant can verify it by comparing expected values against the attested evidence and validating the signature using the platform's public key. 

In \emph{full verification}, the verifier obtains the evidence sets $\mathcal{E}$ along with the signed Merkle root $r$ and reconstructs the tree by computing $r' = \textsc{BuildMerkleTree}(\mathcal{E})$ independently. If $r' = r$ and the signature $\sigma$ over $H(r \,\|\, t \,\|\, \texttt{nonce} \,\|\, \texttt{platform\_id})$ is valid, the report is accepted as authentic and complete. 

For \textit{field-level} verification, the tenant still performs full Merkle root and signature verification as described above but inspects only a specific evidence item $e_i \in \mathcal{E}$. Once the root and signature are validated, the integrity of any individual item, e.g., $e_i$, is implicitly guaranteed by the Merkle commitment. This allows the tenant to selectively check policy-critical fields by verifying a single report.
\label{plaintext}
To reduce communication overhead, only dynamic or runtime-generated fields, such as certain dynamic telemetry subscriptions (see Table~\ref{tab:oru-attestation-scope}), are included in plaintext and sent along with the report,  as tenants may lack prior knowledge of their exact expected value $e_i$, and thus cannot recompute the hash $H(e_i)$ independently. 
In contrast, static configurations (e.g., standardized XML-based NETCONF files) whose hashes are already known to tenants can be omitted, as these configurations are typically designated by the tenants themselves and change infrequently.
 This design enables efficient verification by minimizing data transfer while ensuring that tenants can validate both stable configurations and occasional runtime states as needed.

\noindent \textbf{Response Mechanisms.} 
Upon detecting unauthorized configuration changes, tenants can respond by revoking trust in the compromised O-RU (e.g., via SMO APIs), rerouting traffic to alternative cells, or initiating compliance and forensic workflows. The PRH maintains an immutable audit log of attested reports, ensuring traceable, evidence-backed accountability even in multi-operator or vendor-managed environments. By enabling timely and verifiable responses, the system closes the loop between runtime attestation and operational control.

%% file: 6_eval.tex
\section{Experiment and Evaluation}




\subsection{Experiment Setup and Method} 

We developed a proof-of-concept prototype to evaluate the performance and practicality of our monitoring and verification framework. The experiment involves the deployment of both core components: the TR and the PRH. The TR is responsible for monitoring and attesting critical O-RU evidence items, e.g., NETCONF/YANG configuration states, by generating tamper-evident reports. The reports, along with selective evidence, will be sent to PRH and securely stored by PRH. The verifier can access the reports through PRH and verify their integrity and authenticity. Our prototype was implemented on a desktop equipped with an Intel Core i7-12700K CPU (3.6 GHz, 12-core, 64-bit, 32GB RAM) running Ubuntu 24.04.2 LTS. Each component (TR and PRH) was deployed as a standalone Python application, with communication secured via TLS.

To emulate a realistic O-RU scenario, we deployed the NETCONF server/client using the O-RAN Software Community’s \texttt{sim-o1-ofhmp-interfaces}\footnote{\url{https://github.com/o-ran-sc/sim-o1-ofhmp-interfaces}} repository, which defines standardized YANG models, and XML-based configuration files for O-RU. This simulator provides O1 interface endpoints and open FH behavior, enabling O-RAN-compliant configuration generation and synchronization. It includes operational and startup configuration snapshots commonly used for modeling NETCONF datastore behavior. We implemented the NETCONF server using \texttt{Netopeer2}\footnote{\url{https://github.com/CESNET/netopeer2}}, a production-grade, YANG-driven NETCONF server built atop the \texttt{libnetconf2} and \texttt{sysrepo} libraries. \texttt{Netopeer2} supports dynamic configuration management across standard datastores: \texttt{<running>}, (active configuration), \texttt{<startup>} (persistent configuration loaded at boot), and \texttt{<candidate>} (staging configuration), enabling us to monitor configuration updates and validate consistency across system states.

Both the TR and PRH components were instantiated within a software-based TPM using \texttt{swtpm}\footnote{\url{https://github.com/stefanberger/swtpm}}, a QEMU-compatible TPM 2.0 emulator. TPM functionalities, such as key provisioning, signing, and attestation, were accessed through the official TPM 2.0 Trusted Software Stack, \texttt{tpm2-tss}\footnote{\url{https://github.com/tpm2-software/tpm2-tss}}. We provisioned a 2048-bit RSA Attestation Identity Key (AIK) with the TPM hierarchy using \texttt{tpm2\_createak}, configured to sign the reports using the \texttt{RSASSA-PKCS1-v1\_5} scheme with \texttt{SHA-256} as the hash algorithm. 
The TR used the AIK private key to produce attestation reports, while the PRH and tenant can verify the report using the corresponding AIK public key. Additionally, trust in the AIK is anchored by the TPM’s Endorsement Key (EK), which serves as the hardware-root identity, ensuring the reports originate from a legitimate TPM.


\begin{table}[]
\centering
\caption{Example Evidence Items}\vspace{-5pt}
\resizebox{0.8\columnwidth}{!}{
\begin{tabular}{|p{3.9cm}|p{3.6cm}|}
\hline
\textbf{Attestation Category} & \textbf{Example Config Source} \\
\hline
Startup Configuration Baseline & \texttt{startup-config.xml} \\
\hline
Runtime Operational State & \texttt{running-config.xml} \\
\hline
NETCONF Access Control & \texttt{ietf-netconf-acm.yang} \\
\hline
PTP Synchronization Settings & \texttt{o-ran-sync.yang} \\
\hline
MACsec/IPsec Fronthaul Security & \texttt{o-ran-interfaces.yang} \\
\hline
User Management & \texttt{o-ran-usermgmt.yang} \\
\hline
Hardware Management & \texttt{ietf-hardware.yang} \\
\hline
\end{tabular}
}
\label{tab:attestation-scope-summary}\vspace{-10pt}
\end{table}

\subsection{Report Generation \& Verification}
To evaluate the performance and practicality of our attestation-based monitoring framework, we conducted a series of measurements focused on the latency of report generation and validation. As part of the setup, XML-based configuration files (representative samples are listed in Table~\ref{tab:attestation-scope-summary}) were first loaded into the NETCONF server by the TR. The TR then parses these files to construct a Merkle tree, where each configuration file serves as a leaf node. The \texttt{SHA-256} hash mechanism was applied to each file and subsequently to each pair of child nodes up the tree, resulting in a single 256-bit Merkle root representing a compact cryptographic commitment to the configuration state. 
This Merkle root was then attested using the software-based TPM via a TPM2 quote operation, which generated a 2048-bit RSA signature using the AIK securely stored within the TPM. 
This signature, along with nonce and timestamp metadata, forms the attestation report, which serves as a cryptographic guarantee of the O-RU’s configuration state at a specific point in time.
To assess the effectiveness of the integrity verification mechanism, we introduced adversarial modifications to selected configuration files. During validation, the verifier independently reconstructed the Merkle tree and computed a new root from the modified configuration set. This computed root was then compared against the attested root $r$ received from the TR. A mismatch between the two signaled unauthorized changes, enabling efficient detection of tampering without requiring per-file hash comparisons.

Table~\ref{tab:time} summarizes the timing breakdown of key operations obtained from our prototype, averaged over 10 runs on a configuration involving approximately 50 XML subfiles within the NETCONF directory. Merkle root computation completed in 1.3 ms, reflecting the low computational cost of SHA-256 hashing over a moderate number of inputs and the lightweight nature of leaf-to-root digest aggregation. In contrast, report generation, dominated by the TPM-backed signing operation using a 2048-bit RSA key, incurs a significantly higher latency of 140 ms. This cost stems from the computational expense of modular exponentiation in RSA and the TPM’s I/O overhead for protected key access. On the verifier side, signature verification and root comparison complete in 60 ms. 
The total computation and verification delay for generating and verifying a report was approximately 200 ms, excluding network transmission overhead. 
The attestation payload, including the 256-bit Merkle root, a 2048-bit RSA signature, and metadata, is compact; in cases without plaintext disclosure (e.g., full NETCONF attestation, as discussed in Section \ref{plaintext}), the total size is well under 1 KB.  
Assuming persistent SSH/TLS sessions and typical O-RAN round-trip times (10–100 ms), the overall end-to-end latency aligns with Non-RT RIC timescales, which operate on the order of seconds. As such, our monitoring framework supports timely detection and remediation of misconfigurations or tampering without impacting latency-sensitive RAN functions.

\subsection{Scalability Evaluation}\label{merkle}

\begin{figure}[t]
    \centering
    \includegraphics[width=0.95\linewidth]{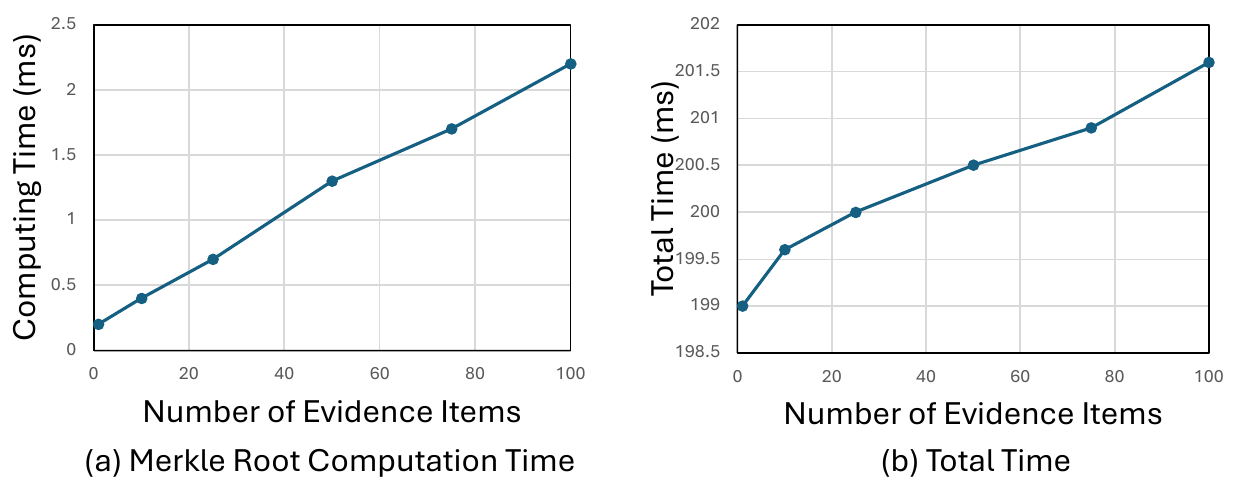}\vspace{-5pt}
    \caption{Merkle Root Computation Time and Total Processing Overhead under Varying Number of Evidence Items}
    \label{fig:time_cost}
    \vspace{-12pt}
\end{figure}

We evaluate the scalability of our monitoring framework by examining its performance, varying the number of evidence items from 1 to 100. Figure~\ref{fig:time_cost} (a) shows the Merkle root computation time at the TR, and Figure~\ref{fig:time_cost} (b) reports the total processing time, which includes Merkle root construction, report generation (i.e., TPM-backed signing), and signature verification at the verifier side.

As shown in Figure~\ref{fig:time_cost} (a), Merkle root construction scales sublinearly with the number of evidence items, staying under 2.5 ms even with 100 items. This aligns with the expected computational cost of SHA-256 hashing over the leaves and intermediate nodes in the tree structure. The low overhead confirms the efficiency of using Merkle trees for aggregating configuration states. Figure~\ref{fig:time_cost} (b) shows that total processing time grows moderately with the number of items, reaching approximately 300 ms at 100 items. This trend is also driven by signing latency during report generation at the TR. In contrast, report verification, comprising Merkle root comparison and signature validation, is comparatively lightweight.

While the figures reflect full recomputation at each step, our framework supports incremental Merkle tree updates by TR, enabling efficient recomputation when only a subset of evidence items changes. This capability provides additional scalability benefits in dynamic environments, where configuration deltas are more common than wholesale changes.

Overall, the monitoring framework remains performant under growing evidence sizes and satisfies the timing requirements for non-real-time RIC operations in O-RAN, making it a practical solution for policy compliance enforcement at scale.

\begin{table}[]
\centering
\small
\caption {Time Consumption}\vspace{-5pt}
\resizebox{0.6\columnwidth}{!}
{
\begin{tabular}{|p{4cm}|p{1.5cm}|}
\hline
Different Stages  & Time ($ms$)\\
  \hline 
   Merkle Root Computation & 1.3\\
   Report Generation at TR & 139.2\\
   Report Verification & 59.6 \\ 
  \hline
  Total & 200.1 \\
  \hline 
\end{tabular}
}
\label{tab:time}\vspace{-10pt}
\end{table}

%% file: 7_discussion.tex
\section{Discussion}
\noindent \textbf{TPM vs. TEE.}
Our core requirement is to protect attestation keys and produce verifiable, cryptographically signed evidence of system state. TPMs fulfill this role effectively, offering secure key storage and platform measurement with minimal operational overhead. As TPMs are increasingly prevalent in modern RAN deployments, this approach provides a practical path to verifiable operations without requiring hardware changes. The TR’s integrity at launch is ensured through TPM-anchored measured boot or Linux Integrity Measurement Architecture (IMA). For deployments requiring stronger protection against runtime tampering, the TR can be hosted within a TEE to provide full in-memory isolation in addition to attestation.

\noindent \textbf{Integrity Assurance vs. Runtime Coverage.}
Our framework provides strong assurance over an O-RU’s intended operational posture by attesting the integrity of its configuration and key runtime states, which govern key functions such as RF behavior, security policies, telemetry, and interface control. This is essential for enforcing tenant policies and maintaining compliance.
However, attesting configurations and selected runtime elements alone cannot fully capture the complexity of live system execution. Persistently recording all runtime states would introduce prohibitive overhead, conflicting with O-RAN’s stringent latency and QoS requirements. Runtime deviations may still arise from edge-case software bugs, hardware failures, or dynamic environmental interactions (e.g., RF variability or C-plane signaling anomalies). While our design establishes a secure and auditable baseline, it can be complemented by anomaly detection or real-time telemetry analytics to capture edge cases and runtime deviations beyond the scope of configuration and operation monitoring.

%% file: 9_conclusion.tex
\section{Conclusions}
Our framework reimagines ZTA in O-RAN by enabling continuous, verifiable oversight of both configuration and runtime control logic. By shifting trust from static credentials to dynamic, cryptographically verified system behavior, it eliminates implicit trust post-authentication and ensures alignment with tenant-defined policies across vendor-diverse deployments.

\section*{Acknowledgment}

This work was supported in part by the Office of Naval
Research under grants N00014-24-1-2730, the US National Science Foundation under grants 2312447, 2247560, 2247561, 2154929, 2332675, 2331936, and 2235232, and the Virginia Commonwealth Cyber Initiative (CCI).